# Peer Influence on Physics Self-Efficacy and Grades: A comparative study of students in an introductory calculus-based course who typically worked alone or in groups before and during the pandemic

Apekshya Ghimire and Chandralekha Singh

*Department of Physics and Astronomy, University of Pittsburgh, 3941 O'Hara St, Pittsburgh, PA, 15260*

Engaging in meaningful collaborations with peers, both inside and outside the classroom, can greatly enhance students' understanding of physics and other STEM disciplines. We analyzed the characteristics of women and men who typically worked alone versus those who collaborated with peers in a calculus-based introductory physics course comparing pre pandemic traditional in-person classes to Zoom based pandemic classes. We discuss our findings by considering students' prior academic preparation, their physics grades and physics self-efficacy, as well as their perceptions of how effective peer collaboration is for their physics self-efficacy. We also compared our results to the first-semester algebra-based introductory physics course.

# I. INTRODUCTION AND FRAMEWORK

Why do some students thrive independently while others shine in group settings? As science teachers, we've all observed these types of differences in our classrooms. Some students work best on their own, while others flourish in collaborative environments where ideas are exchanged and refined. This study dives into how individual versus group learning correlates with students' physics grades and self-efficacy—both in traditional in-person classes and during the remote learning shifts of the COVID-19 pandemic.

Collaborating with peers has long been shown to enhance learning in physics and other STEM disciplines [1-6], boosting students' confidence and performance. Theories such as the distributed cognition framework [7, 8] and the Zone of Proximal Facilitation (ZPF) model [9, 10] highlight how students working together can expand their cognitive capacity, enabling them to tackle complex problems more effectively than when working alone. In STEM education, peer collaboration is especially valuable, as it allows students to reinforce their understanding through discussion, problem-solving, and the exchange of ideas [11, 12].

Collaborative learning can offer numerous benefits, not just for struggling students, but also for high achievers [5]. When students collaborate effectively, their combined strengths lead to better outcomes than when they work alone [13-15]. Think of a study group: one student excels at visualizing problems, another at analyzing data, and a third at making real-world connections. Together, they form a dynamic learning environment that can boost everyone's understanding. But for collaboration to be effective, it must emphasize positive interdependence [16] and individual accountability [17]. Positive interdependence refers to collaborative situations in which all group members rely on each other's strength to achieve their collective goal, fostering cooperation and communication [16]. Individual accountability ensures that every student actively contributes to the group task, maximizing engagement and cognitive effort [17, 18]. This approach not only improves academic performance but also enhances the group's overall productivity, fostering a sense of shared responsibility.

Research shows that peer collaboration can significantly boost self-efficacy [19]—the belief in one's ability to succeed in a task—which is directly linked to improved academic performance [20-24]. This belief is influenced by the classroom environment, teaching strategies, and peer interactions [25, 26], and can significantly impact student engagement, learning, and achievement in science courses [27, 28]. Research shows a positive correlation between higher self-efficacy and academic success [29, 30]. When students collaborate, they not only reinforce their own understanding but also benefit from the perspectives and insights of others, building confidence in their abilities [31, 32]. Modeling effective peer collaboration in class, providing explicit encouragement for collaboration both inside and outside the classroom, and sharing evidence of the benefits of peer collaboration can help increase students' willingness and prioritization of working with peers [33-39]. However, for collaboration to truly enhance learning, it's important to create an inclusive environment where all students, regardless of gender or background, feel valued and encouraged to participate actively [40, 41]. When students believe that their contributions matter, they are more likely to engage in meaningful collaboration.

That said, students face challenges when working together. Issues such as social anxiety, a preference for solo work, or unfamiliarity with the benefits of collaboration can hinder participation [20]. If classroom activities do not emphasize the value of teamwork or model effective peer interactions [42], students might struggle with collaboration. Effective in-class group work can inspire students to engage in collaboration outside of class. Collaboration with peers can also play an important role and influence a student's self-efficacy which is termed as Peer Influence on Self-Efficacy (PISE) (pronounced as "pies"). PISE is a measure of students' perceptions of how interactions with their peers impact their confidence in physics [40]. Furthermore, gender dynamics can influence students' collaborative preferences, especially in STEM fields like physics, where stereotypes about ability can affect confidence and participation. For example, women may feel more comfortable working alone or in same-gender groups due to societal biases, or they might feel marginalized when male peers dominate group discussions [43-47]. Research indicates that women who believe in gender stereotypes in physics tend to perform worse compared to those who do not hold such beliefs [43, 47]. Therefore, it's crucial to create a classroom culture where gender equity is a priority, ensuring that all students feel empowered to share their ideas and collaborate effectively.

The shift to remote learning during the COVID-19 pandemic further impacted students' collaborative experiences, raising important questions about the role of peer interactions in physics self-efficacy and academic success. It brought significant changes to teaching and learning as institutions shifted to remote classes replacing traditional in-person instruction with virtual sessions on platforms like Zoom. Instructors had to develop innovative methods to facilitate student collaboration, employing tools such as breakout rooms, polls, chats, and multimedia presentations. Due to the pandemic's stressful environment, high-stakes exams were discouraged[48], and instructors were advised to use low-stakes assessments throughout the semester. With many students experiencing reduced social interactions with friends and family, virtual platforms provided an opportunity for social engagement and shared experiences. While some students may have managed to do well in this setting, others may have found it challenging. Therefore, it is valuable to evaluate how the teaching approaches as well as how students interact with their course peers during COVID predicts students' academic performance, physics self-efficacy, and their perceptions of the role of peer influence on self-efficacy (PISE).

In this study, as students self-selected to typically work alone or in groups without any incentives, we sought to answer several key research questions. First, we examined whether there are differences in average grades, PISE, and physics self-efficacy between the students who TWA (typically worked alone) and TWG (typically worked in groups) before and during the COVID period. We also explored the proportions of women and men who typically worked alone or in groups. Additionally, we investigated whether any gender differences emerged from these patterns, specifically regarding academic outcomes and physics self-efficacy. Finally, we analyzed whether gender and TWA/TWG predicted PISE and grades, controlling for physics self-efficacy and prior preparation, both before and during the COVID pandemic. To address these research questions, we used a validated survey that assessed PISE, physics self-efficacy, and typical mode of working (group vs. alone) allowing us to examine the relationships between these variables while accounting for gender differences.

Prior research investigated the peer influence on self-efficacy and grades before and during COVID for the first and second algebra-based introductory physics courses in which women outnumber men [49, 50]. In this study, we focused on first calculus-based introductory physics course in which men outnumber women. This course is a part of a two-semester course sequence primarily for students majoring in engineering/physical sciences. We investigated student grades and Peer Influence on Self-Efficacy (PISE) for the students who TWA and TWG and analyzed how these variables were affected by students' gender, physics self-efficacy, and prior preparation. We considered students' interactions with peers both inside and outside the classroom, including collaboration on homework. Since the students decided to typically work in groups or alone without an incentive from the course instructor, this study provides baseline data on the dynamics of students who typically work alone (TWA) versus those who work in groups (TWG) throughout an introductory physics course. This information can serve as a valuable resource for instructors seeking to encourage and model effective peer interactions by integrating evidence-based collaborative practices into their teaching.

## II. RESEARCH QUESTIONS

This study addresses the following research questions for calculus-based introductory physics 1 students:

RQ1. Are there differences in average grades, Peer Influence on Self-Efficacy (PISE), and physics self-efficacy between students who typically worked alone and those who typically worked in groups before and during the COVID pandemic?

RQ2. What percentages of women and men typically worked alone or in groups before and during the COVID pandemic?

RQ3. Do the patterns observed in RQ1 reveal any gender differences in grades, PISE and physics self-efficacy between students who TWA and TWG for each gender?

RQ4. Do gender and TWA/TWG predict PISE and grades controlling for physics self-efficacy and prior preparation, before and during the pandemic?

## III. METHODOLOGY

### 2.1 Participants

The study was conducted at a large U.S. public research university involving students from a calculus-based introductory Physics 1 course. The survey was administered during three consecutive fall semesters, spanning both the pre-pandemic period and the time of the COVID-19 pandemic. 502 students completed the survey before the pandemic, and 520 students completed the survey during the pandemic when classes were held synchronously via Zoom. The survey was given during the final week of mandatory teaching assistant-led recitations. The survey asked students whether they typically worked alone or in groups. It also included questions related to physics self-efficacy and peer influence on self-efficacy.

## 2.2 Prior Academic Performance

Prior academic preparation was evaluated using high school grade point averages (HS GPAs) and Scholastic Assessment Test (SAT) scores. HS GPAs were reported on a 5.0 scale, with students having GPAs above 5.0 excluded from analysis due to potential grading system differences. SAT scores [51] were used along with converted American College Testing (ACT) scores, utilizing established conversion tables[52]. We only used the SAT Math and ACT Math scores for our study. Both demographic and prior academic data were obtained from de-identified university records via an honest broker, ensuring confidentiality and preventing knowledge of participants' identities.

## 2.3 Physics Grade

Final course grades were used as academic performance metrics, based on students' performance in homework, quizzes, midterms, and final exams. The final grade, representing overall performance, was linked to students' gender and survey responses. Grades were reported on a 0-4 scale, with A = 4, B = 3, C = 2, D = 1, F = 0, or W (late withdrawal). Grade modifiers '+' and '-' adjusted grade points by 0.25 (e.g., B- = 2.75 and B+ = 3.25), with A+ reported as 4.

## 2.4 Survey

Table 1 shows Peer influence on self-efficacy (PISE) and Physics self-efficacy (SE) [19](adapted from previously validated surveys [44, 53-55] and Peer influence on self-efficacy (PISE) [5] items designed on a Likert scale of 1-4 [56] with 1 – Strongly Disagree, 2 – Disagree, 3- Agree, 4 – Strongly Agree. To ensure that we were measuring domain-specific psychological constructs, we explicitly mentioned physics in the survey items.

Table I: The peer influence on self-efficacy (PISE) and physics self-efficacy items on the survey.

| Item No. | My experiences and interactions with other students in this class: |
|---|---|
| PISE 1 | Made me feel more relaxed about learning physics |
| PISE 2 | Increased my confidence in my ability to do physics |
| PISE 3 | Increased my confidence that I can succeed in physics |
| PISE 4 | Increased my confidence in my ability to handle difficult physics problems |
| SE 1 | I am able to help my classmates with physics in the laboratory or in recitation. |

| SE 2 | I understand concepts I have studied in physics. |
| SE 3 | If I study, I will do well on a physics test. |
| SE 4 | If I encounter a setback in a physics exam, I can overcome it. |

The instructors did not provide any incentives for students to collaborate with one another. Consequently, students chose to collaborate on their own accord, and their responses are self-reported based on interactions both within and outside the classroom setting pertaining to the course for whether they typically worked alone (TWA) or typically worked in groups (TWG). The term "typically" refers to their primary mode of working, either alone or in groups.

**2.5 Analysis**

We standardized PISE, physics self-efficacy, and student grades. Each student's average for the four PISE-related items (see Table I) was calculated and converted to z-scores using the formula $Z = \frac{X - \bar{X}}{\sigma_X}$. This process converted observations into standard deviations from the mean and these were rescaled to a range of 0-1 using the formula $(Z - Z_{MIN})/(Z - Z_{MAX})$[57], which was done separately for non-COVID and COVID semesters. This standardization process was also applied to grades and physics self-efficacy items.

To analyze differences between students who TWA and TWG, we conducted unpaired *t*-tests and calculated Cohen's *d* for effect size. Note that while gender is not a binary construct, the university data only included binary categories of women and men. Cohen's *d* was computed using $d = \frac{\bar{X}_1 - \bar{X}_2}{S_{pooled}}$, where $\bar{X}_1$ and $\bar{X}_2$ are the sample means of the two groups and $S_{pooled}$ is the pooled standard deviation [57]. Similar to common practice [57], we defined small $d \sim 0.2$, medium $d \sim 0.5$, and large $d \sim 0.8$ with statistical significance indicated by $p < 0.05$.

We used multiple regression analysis to explore how PISE and grades are predicted by various factors, including gender, HS GPA, SAT Math scores, and physics self-efficacy. For each regression model, we reported standardized β coefficients, sample size, and Adjusted R-squared. Adjusted R-squared accounts for the number of predictors, making it suitable for models with multiple variables. Standardized coefficients facilitate comparison in terms of standard deviations [57].

**IV. RESULTS AND DISCUSSION**

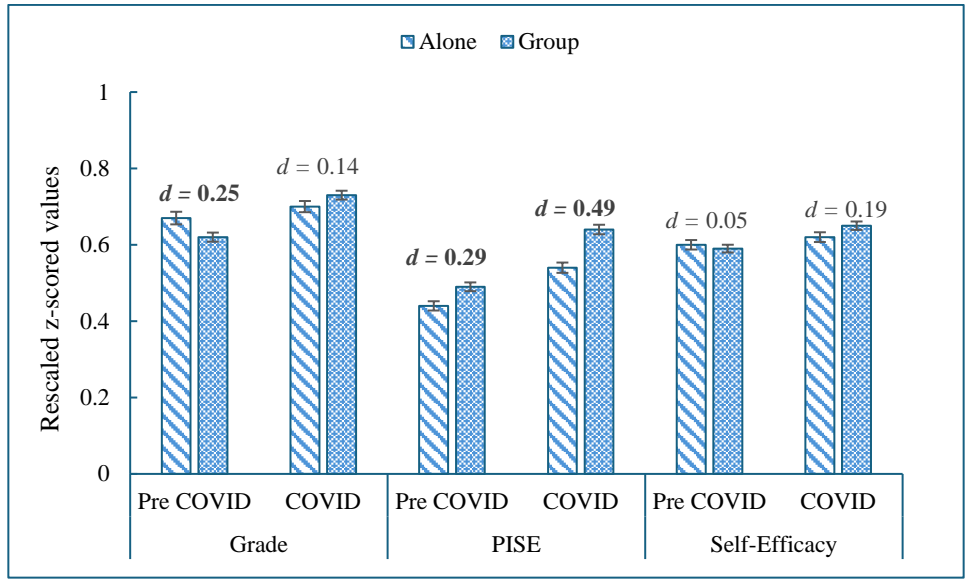

Fig. 1: Grade, PISE and SE of students who typically worked alone and in groups (for both Pre COVID and COVID) with effect size given by Cohen's *d*.

In our examination of disparities in academic performance, peer influence on self-efficacy (PISE), and physics self-efficacy between students who TWA versus those who TWG, we present data without gender separation initially. Prior to COVID, students who worked in groups (TWG) had significantly lower grades than those who worked alone (TWA), as shown in **bold** in Fig. 1. But during the pandemic, that gap pretty much disappeared—the grade difference between the two groups was small and not statistically significant, meaning it didn't have much real impact. However, when it came to Peer Influence on Self-Efficacy (PISE), the difference between TWA and TWG students stuck around. In fact, the gap actually grew, with the effect size increasing from 0.29 to 0.49. Regarding physics self-efficacy, our analysis found no statistically significant differences between these two groups, both before and during the pandemic suggesting that the mode of collaboration did not substantially influence students' self-perception of their abilities in either context.

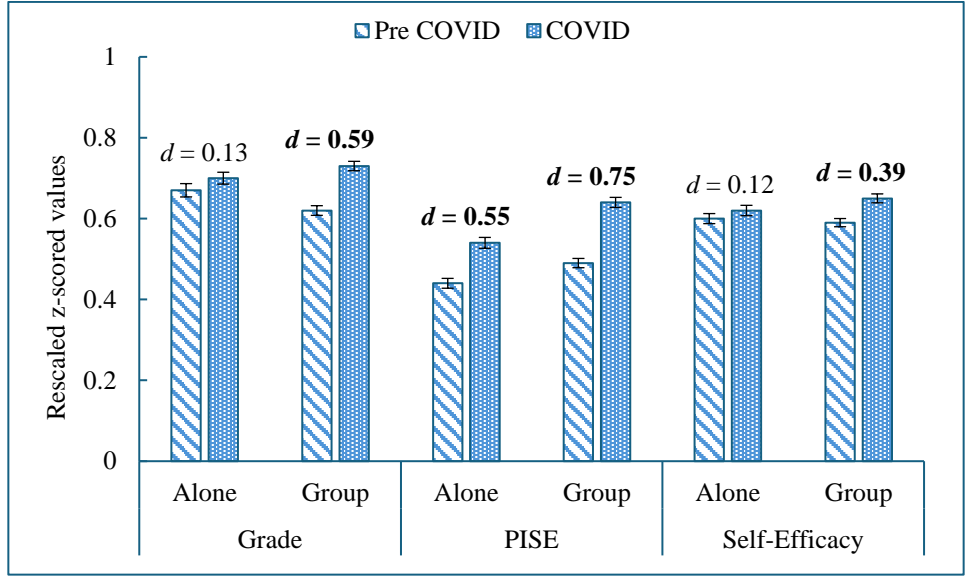

Fig. 2: Grade, PISE and Self-efficacy before and during COVID of the students who TWA and those who TWG.

We also looked at how TWA and TWG students compared before and during COVID, as shown in Fig. 2. For students who TWG, we found a significant difference. Those who collaborated during COVID ended up with higher grades, PISE, and physics self-efficacy than their pre-COVID peers. But for students who TWA, the only significant difference was in PISE with higher PISE among COVID students. However, their grades and physics self-efficacy didn't change much between the two time periods.

For RQ2, we investigated the percentage of students who TWA or TWG before and during the COVID pandemic. As detailed in Table II, about two thirds of the women and half of the men TWG before the COVID-19 pandemic. Before the pandemic, a greater percentage of women (62%) worked in groups compared to men (51%). During the pandemic, the percentage of women working in groups increased to 66% while the percentage of men decreased to 46%.

Table II: Number of women and men who worked alone and in groups before and during COVID.

|  | Before COVID | | During COVID | |
|---|---|---|---|---|
|  | Women | Men | Women | Men |
| Alone | 69 (38%) | 156 (49%) | 72 (34%) | 166 (54%) |
| Group | 111 (62%) | 166 (51%) | 138 (66%) | 144 (46%) |
| $\chi^2$ | 4.78 (p = 0.029) | | 18.72 (p < 0.001) | |

Since we did not collect students' reasons for choosing to TWA or TWG, the cause of this differing trend between women and men remains unclear. Various factors may have influenced their choices before and during the COVID pandemic. While a lower percentage of women TWA in both periods, the reasons for doing so may have differed. For some students, challenges such as household responsibilities like taking care of younger siblings, helping with family duties, or not having a private space for studying could have made it more difficult to collaborate effectively over Zoom. These factors, along with scheduling conflicts, may have influenced students' decisions to work alone during the pandemic.

Regarding RQ3, as illustrated in Fig. 3, women received lower grades compared to men. However, there was no statistically significant difference in performance by gender for students who TWA or TWG both before and during the pandemic, with all groups showing a small effect size (Cohen's $d \leq 0.20$). It appears that gender did not play a significant role in determining the course grades for the students in this calculus-based introductory course.

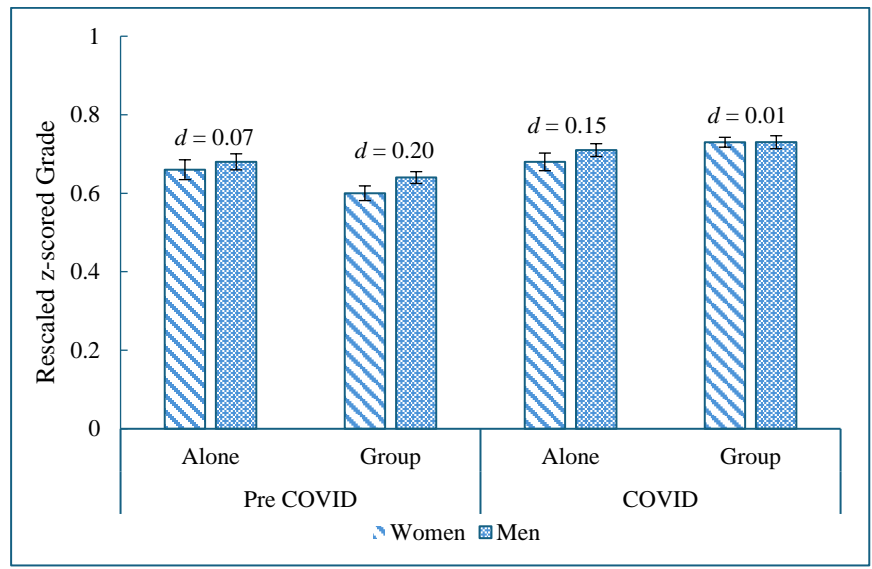

Fig. 3: Grades of women and men who TWA and TWG before and during COVID.

Before the pandemic, women who TWA actually had significantly higher grades than those who worked in groups (TWG), with a medium effect size of 0.33 (see Fig. 4). During the COVID-19 period, this trend reversed, with women scoring higher grades when working with others, with an effect size of 0.29. It will be useful to conduct interviews with students to understand whether the shift to remote learning provided a consistent experience for women and men and how, e.g., issues related to stereotype threat were impacted during remote learning. For men, the grade differences between those who TWA or TWG remained statistically insignificant throughout.

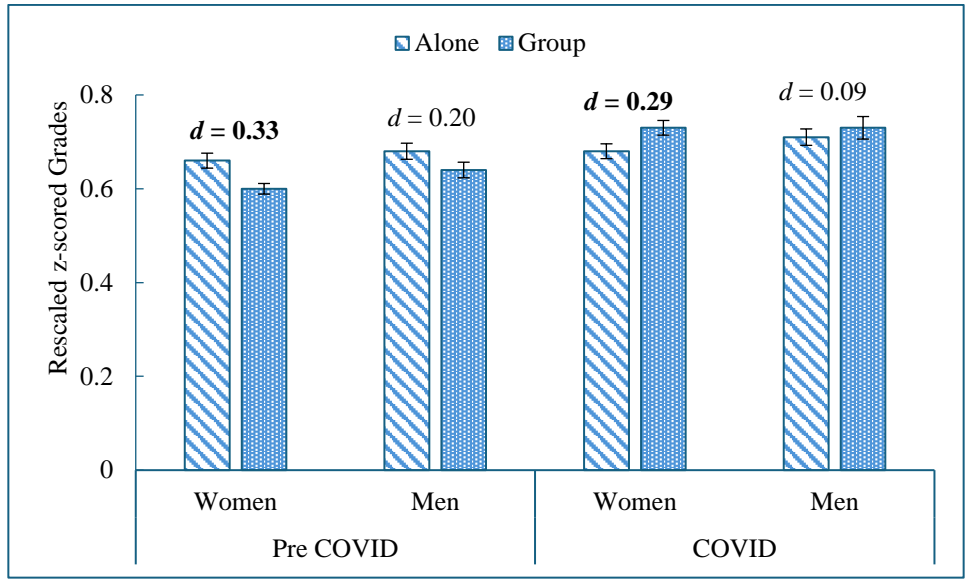

Fig. 4: Grades of the students who TWA/TWG for women and men before and during COVID.

The gender difference in PISE was statistically significant across all groups of students who TWA and TWG both before and during COVID with higher PISE among men compared to women (see Fig. 5). Before COVID, the gender difference was almost comparable, whereas during the pandemic, it was more pronounced among students who TWA.

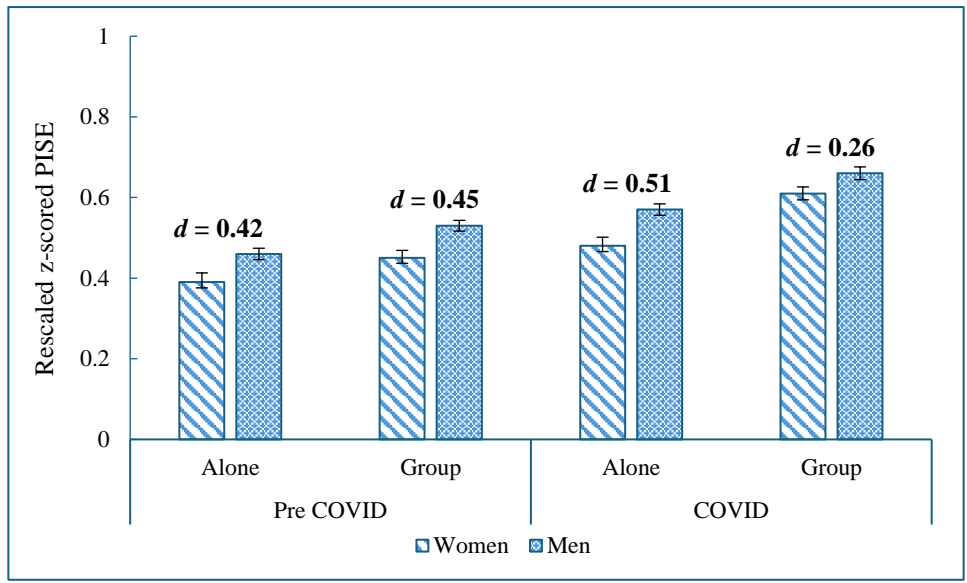

Fig. 5: PISE of women and men who TWA and TWG before and during COVID.

When we looked at PISE scores for students who TWA versus those who TWG within each gender group (see Fig. 6), we see some interesting patterns. Before the pandemic, there wasn't much of a difference for women whether they TWA or TWG, their scores were relatively similar.

But during COVID, that changed in a big way. Women who TWG saw a major boost in their PISE, with a relatively large effect size ($d \sim 0.71$), meaning group work really made a difference.

For men, the gap between the students who TWA and TWG was already there before COVID, and it stayed statistically significant throughout. That said, Fig. 6 shows that the difference became even more noticeable during the pandemic, with an effect size of 0.48, showing that working in groups continued to be beneficial.

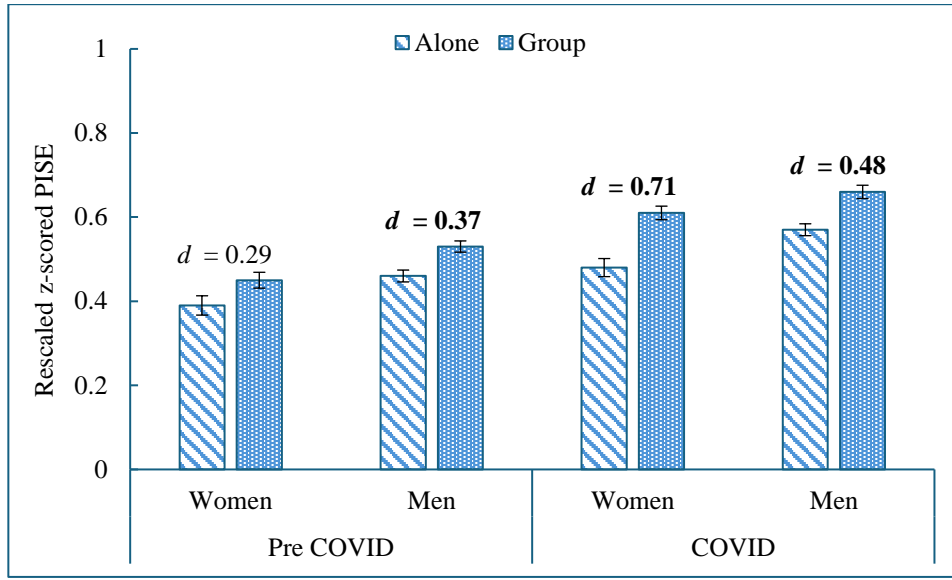

Fig. 6: PISE of the students who TWA/TWG for women and men before and during COVID.

We found clear gender differences in physics self-efficacy across the board—whether students TWA or TWG, both before and during COVID. Men consistently reported higher self-efficacy than women (see Fig. 7) consistent with some prior research studies, e.g.,[58].

Before the pandemic, this gender gap was relatively steady in both groups. But during COVID, things shifted. The gap widened even more for students who TWA, jumping to an effect size of $d \sim 0.66$, while for those who TWG, it remained slightly smaller at $d \sim 0.52$. This suggests that working alone during the pandemic may have potentially amplified existing disparities.

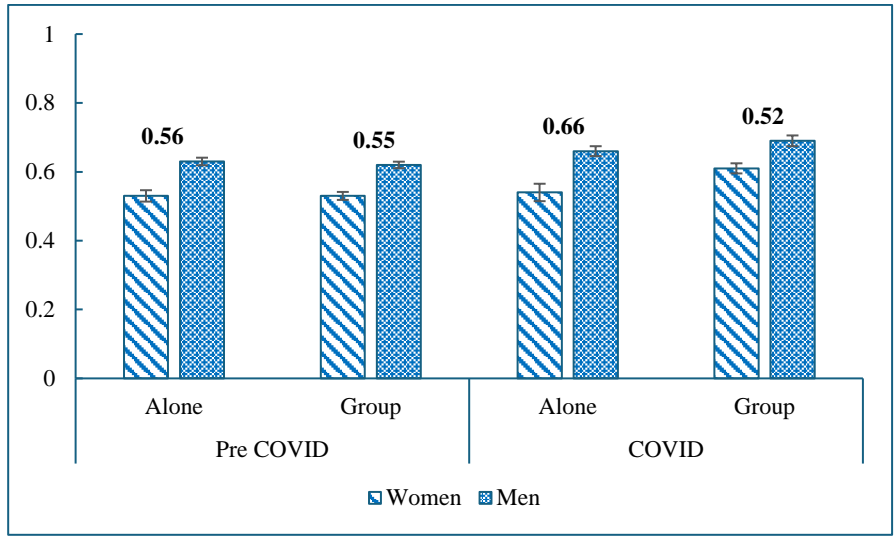

Fig. 7: Physics Self-Efficacy of women and men who TWA and TWG before and during COVID.

When we looked at physics self-efficacy between students who TWA and those who TWG for each gender (see Fig. 8), there wasn't much of a difference before the pandemic—for both women and men, working alone or in groups didn't seem to correlate with their confidence levels.

But during COVID, we saw that women who TWG saw a boost in their physics self-efficacy, with a solid effect size of $d \sim 0.43$, meaning group work really made a difference. For men, the gap also became significant during the pandemic, though the effect was smaller ($d \sim 0.23$), suggesting that while group work appears to have helped both genders in this regard, women seemed to be benefitting more from it.

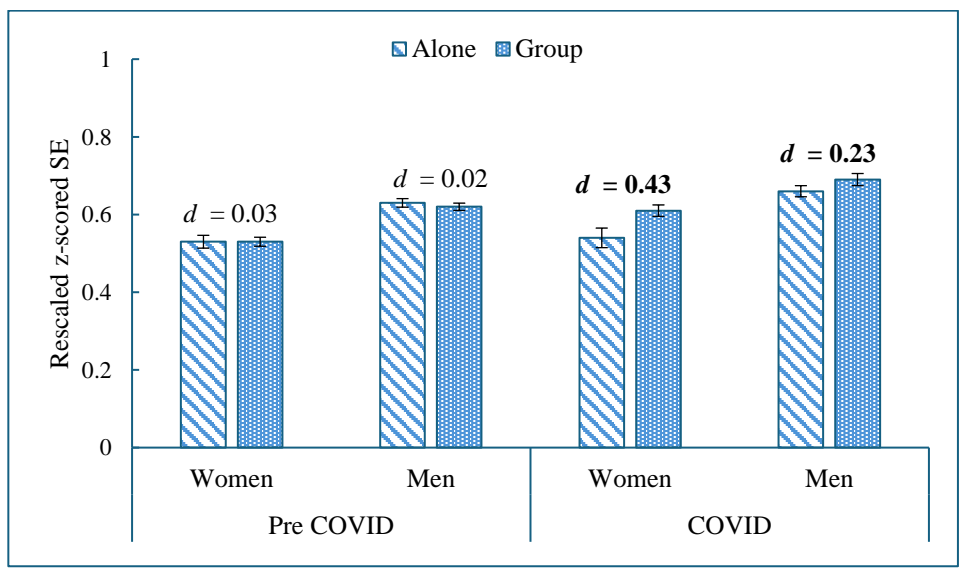

Fig. 8: Self-Efficacy of the students who TWA/TWG for women and men before and during COVID.

Regarding RQ4, Table III shows the results of some useful regression models predicting course grade for calculus-based physics. The predictors (independent variables) used in these models are listed in the first column, with blank spaces indicating predictors not included in a specific model. The effectiveness of each predictor is represented by the standardized coefficients (β), accounting for all other variables. This means that for each standard deviation increase in a predictor variable, the model predicts a corresponding β standard deviation change in the outcome variable, while controlling for other predictors. Although multiple regression models were developed for grades as outcome, we highlight only those models that exhibit significant differences and hold the most relevance.

Table III: Standardized coefficients of regression models predicting grades. SE = Physics Self-Efficacy; Alone/Group = Students who TWA/TWG. Significant predictors are bold. * = $p < 0.05$, ** = $p < 0.01$, and *** = $p < 0.001$[57].

| **Before COVID (Grade as the outcome variable)** | | | | |
|---|---|---|---|---|
| Predictors | Model 1 | Model 2 | Model 3 | Model 4 |
| Gender | 0.07 | -0.05 | 0.10* | 0.01 |
| Alone/Group | **-0.12**** | **-0.12**** | **-0.10*** | **-0.11**** |
| HS GPA | | | **0.28**** | **0.27**** |
| SAT Math | | | **0.38**** | **0.32**** |
| SE | | **0.43**** | | **0.31**** |
| Adj. $R^2$ | 0.02 | 0.19 | 0.27 | 0.40 |
| **During COVID (Grade as the outcome variable)** | | | | |
| Predictors | Model 1 | Model 2 | Model 3 | Model 4 |
| Gender | 0.03 | -0.05 | 0.04 | -0.02 |
| Alone/Group | 0.08 | 0.03 | 0.06 | 0.02 |
| HS GPA | | | **0.21**** | **0.22**** |
| SAT Math | | | **0.45**** | **0.39**** |
| SE | | **0.33**** | | **0.24**** |
| Adj. $R^2$ | 0.01 | 0.10 | 0.26 | 0.31 |

Looking at the regression models for grades before COVID (Table III), gender doesn't seem to matter—men and women performed about the same in this course. But whether students typically worked alone or in groups did make a difference. Across all models, those students who TWA tended to score higher than those who TWG, as shown by the negative coefficient for the Alone/Group variable. Model 1, which only looks at gender and working mode, doesn't explain much of the grade differences (low adjusted $R^2$). Things get more interesting in Model 2, where physics self-efficacy (SE) is added as a predictor. SE turns out to be statistically significant, and the adjusted $R^2$ jumps to 0.19—suggesting that students who have higher physics self-efficacy tend to score higher in this course.

Model 3 brings in prior preparation factors like high school GPA and SAT Math scores, and these also seem to be strong predictors of course grades. Students with stronger academic backgrounds generally earn better grades. Finally, Model 4, which includes both prior preparation and self-efficacy, does the best job of explaining grade differences. This suggests that a mix of solid earlier preparation and confidence in physics is the best predictor of success in the course.

During COVID, neither gender nor whether students TWA or TWG seem to make a difference in how students performed. What did matter, though, was prior preparation (high school GPA and SAT Math scores) along with physics self-efficacy. These remained the strongest predictors of grades, with Model 4 showing an adjusted $R^2$ of 0.31, meaning that students with a solid academic background and confidence in physics tended to do better in the course.

Table IV: Standardized coefficients of regression models predicting PISE. SE = Physics Self-Efficacy; Alone/Group = Students who TWA/TWG. Significant predictors are bold. (* = $p < 0.05$, ** = $p < 0.01$, and *** = $p < 0.001$)[57].

| | **Before COVID (PISE as the outcome variable)** | | | |
|---|---|---|---|---|
| Predictors | Model 1 | Model 2 | Model 3 | Model 4 |
| Gender | **0.19***  | **0.20*** | | 0.07 |
| Alone/Group | | **0.16*** | | **0.17*** |
| SE | | | **0.53*** | **0.52*** |
| Adj. $R^2$ | 0.03 | 0.06 | 0.28 | 0.31 |
| | **During COVID (PISE as the outcome variable)** | | | |
| Predictors | Model 1 | Model 2 | Model 3 | Model 4 |
| Gender | **0.12** | **0.17*** | | 0.05 |
| Alone/Group | | **0.27*** | | **0.21*** |
| SE | | | **0.47*** | **0.44*** |
| Adj. $R^2$ | 0.01 | 0.08 | 0.22 | 0.26 |

Before COVID, Model 1 shows that gender played a big role in predicting PISE, with men reporting higher PISE than women as shown in Table IV. But when we factor in whether students TWA or TWG, Models 2 and 4 reveal that Alone/Group is a strong predictor of higher PISE. In Model 3, physics self-efficacy alone explains 28% of the variation in PISE, showing that students who feel more confident in physics tend to have higher PISE. Model 4, which includes gender, working style, and self-efficacy, explains the most variance (adjusted $R^2$ = 0.31). Interestingly, once we account for working style and self-efficacy, gender no longer plays a significant role in predicting PISE.

During COVID, we see the same patterns across all models (Table IV), with Model 4 still explaining the most variance. This suggests that whether classes were in-person or remote didn't change how gender, working mode, or self-efficacy correlated to PISE in this physics course.

## V. CONCLUSION AND TEACHING IMPLICATIONS

Our study offers baseline data from a time when there was no incentive for students to engage in group work, making these findings useful for instructors and others researching the impact of peer collaboration in more controlled environments.

For RQ1, before the pandemic, students who TWA earned significantly higher grades than those who TWG. But during the pandemic, the trend flipped—TWG students actually outperformed TWA students, though the difference wasn't statistically significant. Interestingly, students who collaborated during the pandemic did much better than those who worked in groups pre-pandemic. This suggests that remote group work may have become more effective, possibly because it served as both an academic and social support system during a time of isolation.

In terms of PISE, TWG students consistently reported higher PISE than TWA students, both before and during the pandemic. Also, PISE increased for everyone during remote learning, regardless of whether they worked alone or in groups. However, their physics self-efficacy didn't show a big difference between the two groups in either time period.

From RQ2, we found that two-thirds of women TWG and one-third TWA across both time periods. For men, the split was roughly even before the pandemic, but during the pandemic, the proportion shifted, with 54% who TWA and 46% who TWG, indicating a shift towards working alone.

Regarding RQ3, we didn't find any major gender differences in grades among students who TWA or TWG across all time periods. However, there were statistically significant gender differences in PISE and self-efficacy within the students who TWA and TWG in both time periods. Women who TWA actually earned higher grades than those who TWG before the pandemic. But during the pandemic, the pattern flipped and women who collaborated in groups did better. For men, there was no real difference in grades among men who TWA and TWG both before and during the pandemic.

Also, men who TWG had higher PISE than those who TWA both before and during the pandemic. This difference was statistically significant only during the pandemic for women suggesting that virtual collaboration may have created a more comfortable environment for some students to engage with their peers. The online format might have reduced some of the social pressures that can make in-person group work challenging for certain students. In terms of self-efficacy, students who TWG reported significantly higher confidence in their physics abilities than those who TWA, and this was true for both men and women during the pandemic.

Regarding RQ4, during the pandemic, neither gender nor working style (TWA vs. TWG) seemed to make a difference. But before the pandemic, students who TWA tended to earn higher grades. The strongest predictors of course performance was prior academic preparation—things like high school GPA and SAT Math scores—as well as physics self-efficacy. Finally, self-efficacy played the biggest role in shaping PISE— students with higher self-efficacy (greater confidence in their physics abilities) were more likely to interact with peers in ways that reinforced and further enhanced their self-efficacy. But gender and working styles also mattered, especially when self-efficacy wasn't factored in.

When comparing our results to those from first-semester algebra-based introductory physics course, we observe a key difference in RQ1: students who TWG obtained significantly higher grades than those who TWA during the pandemic in the first algebra-based course [49]. However, this difference is not statistically significant in the calculus-based course. In fact, students who TWA earned significantly higher grades than those who TWG before the pandemic. Regarding RQ2, we found that the percentages of women and men who TWA and TWG are quite similar in both courses. Also, the shift observed during COVID, with an increase in the percentage of men who TWA, was consistent across both courses. Regarding RQ3, there were no gender differences in grades for the calculus-based course. However, significant gender differences were observed only among students who TWG before COVID in the algebra-based course. There were significant gender differences in PISE for all groups in the calculus-based course, but only among those who TWG before COVID in the algebra-based course. For self-efficacy, we observed significant gender differences across all groups and time periods consistent with some prior research [58]. We also found that prior preparation and self-efficacy are the strongest predictors of grades and PISE in both algebra-based and calculus-based courses.

Since this study did not involve interview data, we do not know the reasons for why students typically worked alone or in groups before or during the pandemic. The pandemic-imposed isolation on many students through lockdowns and social distancing, may have potentially affected their academic and social lives. To better understand student behavior and its influence on learning, in-depth interviews can be invaluable and can shed light on the advantages of peer collaboration in both online and traditional classroom settings. These one-on-one interviews could delve into students' perceptions of social interaction and peer collaboration, whether on Zoom during the pandemic or in-person classes. They could explore gender-specific experiences, differences in physics self-efficacy and academic performance, as well as the technological and logistical challenges in online and in-person collaboration.

Overall, our findings highlight the relationship of peer collaboration with gender, self-efficacy and academic performance in physics education. While students who worked alone (TWA) performed better before the pandemic, group work (TWG) became more effective during social isolation. We also observed that a higher percentage of women typically worked in groups during the pandemic and more men typically worked in groups before the pandemic. Although grades did not significantly differ by gender, we observed that women who TWA got statistically significantly higher grades before COVID whereas during the pandemic, women who TWG got statistically significantly higher grades. We did not see any significant differences between these two groups for men in both the time periods. Gender disparities emerged in PISE and self-efficacy, where men always had a higher PISE and a higher self-efficacy than women across both groups and time periods. The gender differences in PISE and self-efficacy were much larger for students who TWA during COVID compared to other groups. We also observed that women who TWG had a significantly higher PISE and self-efficacy than those who TWA during the pandemic. Prior academic preparation and self-efficacy emerged as key predictors of academic success, with self-efficacy emerging as the strongest predictor of PISE. This emphasizes the importance of creating learning environments that foster confidence and preparedness for all students in physics learning.

These insights suggest that encouraging inclusive and supportive group work environments, while also addressing individual self-efficacy and preparation, could enhance student outcomes in physics education. Science teachers might consider offering flexible options for both group and individual work, recognizing that different students thrive in different settings. Virtual collaboration tools could be valuable even in face-to-face classes, as they might provide an additional layer of comfort for some students. Teachers can create structured opportunities for peer support that don't require students to initiate collaboration themselves, while remaining mindful of group dynamics that might affect underrepresented students. Beginning with low-stakes group activities can help build student confidence before tackling more challenging problems.

By understanding these patterns in how students work together and learn, we can better support all our students. The key lies in creating flexible learning environments that honor both individual and group work preferences while ensuring everyone has access to the benefits of peer collaboration when they need it. This balanced approach can help create more inclusive and effective classrooms where all students have the opportunity to succeed.

## DATA AVAILABILITY STATEMENT

The data presented in this study are available upon reasonable request from the corresponding author due to data privacy requirements of US FERPA regulations.